\documentclass[twocolumn,showpacs,prl,aps,tightenlines,superscriptaddress]{revtex4}

\usepackage{graphics,graphicx}

\newcommand{\beq}{\begin{equation}}
\newcommand{\eeq}{\end{equation}}
\newcommand{\bea}{\begin{eqnarray}}
\newcommand{\eea}{\end{eqnarray}}

\newcommand{\lie}{\mathcal{L}}

\begin{document}

\title{	
Geometry and Regularity of Moving Punctures
}

\author{Mark Hannam} \affiliation{Theoretical Physics Institute, University of Jena, 07743 Jena, Germany}
\author{Sascha Husa} \affiliation{Theoretical Physics Institute, University of Jena, 07743 Jena, Germany}
\author{Denis Pollney} \affiliation{Max-Planck-Institut f\"ur Gravitationsphysik, Albert-Einstein-Institut, Am M\"uhlenberg 1, 14476 Golm, Germany}
\author{Bernd Br\"ugmann} \affiliation{Theoretical Physics Institute, University of Jena, 07743 Jena, Germany}
\author{Niall \'O~Murchadha} \affiliation{Physics Department, University College Cork, Ireland}

\date{June 16, 2006; 
      July 25, 2007}

\begin{abstract}
Significant advances in numerical simulations of black-hole
binaries have recently been achieved using the puncture method. We examine
how and why this method works by evolving a single black hole. The
coordinate singularity and hence the geometry at the puncture are found to
change during evolution, from representing an asymptotically flat end to
being a cylinder.  We construct an analytic solution for the
stationary state of a black hole in spherical symmetry that matches
the numerical result and demonstrates that the evolution
is not dominated by artefacts at the puncture but indeed finds the
analytical result.
\end{abstract}

\pacs{
04.20.Ex,   
04.25.Dm, 
04.30.Db, 
95.30.Sf    
}

\maketitle
Recent breakthroughs in numerical relativity have made it possible to simulate
the evolution of black-hole binaries through several orbits, inspiral, merger
and ringdown
\cite{Campanelli:2005dd,Baker05a,Pretorius:2005gq,Bruegmann:2003aw}. 
It is now possible, after over forty years of research, to study
black-hole mergers in full general relativity, and to calculate the resulting
gravitational-radiation waveforms. These advances represent a major
step forward for all of black-hole and gravitational-wave astronomy and
astrophysics. 

Although these methods have met with spectacular success, it is not yet clear
how and why they work. In this letter we focus on the most popular method,
called ``moving punctures'' \cite{Campanelli:2005dd,Baker05a}, and use simple
geometrical considerations to address this question. 
In the moving-puncture method the black holes are 
conveniently described in the initial data in coordinates that do not reach the 
black holes' physical singularities; as the coordinates approach each 
singularity they instead follow a wormhole through to another copy of the 
asymptotically flat exterior spacetime. These extra copies 
are compactified so that their infinities are represented by single points
on a numerical grid, which are called ``punctures''
\cite{Brill63,Brandt97b}. In this construction,  
the wormhole topology is captured by a single function, a conformal factor
$\psi$, which diverges at each puncture. Early ``fixed puncture'' evolutions
\cite{Bruegmann97,Alcubierre00b}
factored the singularity into an analytically prescribed conformal
factor, and the gauge conditions prohibited the punctures from moving
across the grid. 
The ``moving puncture'' approach manages to evolve the full conformal
factor, $\psi$, and the punctures are allowed to move. This seemingly minor
modification proved to be the last piece of the black-hole binary puzzle, and
has made long-term stable simulations routine for many research groups.

However, the dynamical behavior of the punctures in this method 
is entirely unknown. Do they continue to represent compactified infinities?
Does the evolution reach a final, stationary state, or do gauge dynamics
persist? And, crucially, does the method accurately describe the spacetime, or
does it rely on numerical errors near an under-resolved puncture, implying that 
it may fail when probed at higher resolutions or for longer evolutions? 
Finally, why do our coordinate conditions accurately reproduce
angular velocities as would be measured in an asymptotic rest frame 
(see e.g.\ \cite{Diener:2005mg})?

In this letter, we address these questions in three stages. First, we
argue that the evolution of the punctures can be split approximately
into advection across the hypersurface plus a gauge that locally
produces black holes in a stationary state. Second, we assume
existence of a stationary solution and show that a local expansion for
a Schwarzschild puncture matches the numerical data.  And third, we
explicitly construct stationary slices for Schwarzschild and discuss
their global geometry, to which the numerical evolution asymptotes.

Most significantly, we find that the puncture changes character: the slice no
longer approaches the other asymptotically flat end, but ends on a cylinder of
finite areal radius. This suggests an elegant new way to represent black
holes, with asymptotically cylindrical data.  

Let $r$ be the distance to one of the punctures. Internal
asymptotically flat ends are characterized by 3-metrics of the form
$g_{ij} = \psi^4 \tilde g_{ij}$, where $\tilde g_{ij}$ is finite and
the conformal factor $\psi$ diverges as $\psi
\sim 1/r$.  The method of~\cite{Campanelli:2005dd} introduces a
regular conformal factor $\chi=\psi^{-4}$, 
while \cite{Baker05a} uses $\phi=\log\psi$. Ignoring
the $\log r$ singularity, standard finite-differencing is used near the
punctures. 
The evolution method is based on the
Baumgarte-Shapiro-Shibata-Nakamura (BSSN) formulation \cite{Shibata95,Baumgarte99}.
For the lapse we consider ``1+log'' slicing, 
$(\partial_t - \beta^i\partial_i) \alpha = -2\alpha K,$
and for the shift the ``Gamma-freezing'' condition, 
$\partial_t^2 \beta^i = \frac{3}{4} \partial_t \tilde \Gamma^i 
	- \eta \partial_t\beta^i$
\cite{Alcubierre02a,Campanelli:2005dd,vanMeter:2006vi,Gundlach:2006tw}. 

Puncture evolutions use the standard 3+1 decomposition, and the
equations can be brought into the form
	$(\partial_t - \lie_\beta) u = F$,
where $u$ is a state vector and $F$ a source term independent of the
shift. In this context we do not have to consider the BSSN variable
$\tilde\Gamma^i$ since it is derived from $\tilde g_{ij}$. 
Of special importance is the motion of the puncture itself,
which is marked by a $\log r$ pole in $\phi$ or equivalently a zero in
$\chi$~\cite{Campanelli:2005dd}, which are advected by the shift like all
other variables.
As noted in~\cite{Bruegmann97,Alcubierre02a,Campanelli:2005dd,Baker05a}, since
the shift can describe arbitrary coordinate motion, any shift vector that does
not vanish at the puncture can generate motion of the puncture.

The Gamma-freezing shift 
manages two feats, stabilizing the black holes
against slice stretching, and moving the punctures. 
Considering that in
actual runs those two effects appear to happen independently of each
other, we propose
that the shift can be split as
$\beta^i = \beta^i_{adv} + \beta^i_{sls},$
where $\beta^i_{sls}$ counters slice-stretching and $\beta^i_{adv}$ is
responsible for puncture motion. 
If we have {\em exact} stationarity, then
    $(\partial_t - \lie_{\beta_{adv}}) u = 0$ and
	$\lie_{\beta_{sls}} u + F = 0$.
Note that in the early phase of a binary inspiral
simulation the vector $(\frac{\partial}{\partial t})^a$ should be an approximate helical
Killing vector, up to a rigid rotation. We can choose either corotation,
or vanishing rotation at infinity, which would imply that the punctures' coordinate 
speeds equal their physical speeds as seen from infinity.

In general, 
this shift decomposition will hold 
only approximately, but it should be
an excellent approximation in the immediate vicinity of the punctures.
Furthermore, if the shift is regular at a puncture with a leading
constant plus higher order terms,
	$\beta^i = b_0^i + O(r)$, 
$\left.\beta_{adv}^i\right|_{r=0} = b_0^i$, 
$\left.\beta_{sls}^i\right|_{r=0} = 0$,
then we obtain trivial advection of the puncture without
further approximations, i.e., $u=u(x+b_0 t)$, 
since $\lie_{b_0} u= b_0^i\partial_i u$.

We conclude that, assuming approximate stationarity, the
question of numerical stability of the moving puncture process can be approached
by splitting it into a standard advection problem 
and a stability analysis of a stationary puncture solution. 
The main open issue is whether there exists a regular,
stationary solution for a single puncture that the method can find.
Since the key novel features of this stationary solution already occur
for a non-moving, spherically symmetric puncture (computed with the
moving puncture method), we focus on this case here and leave the
general case to future work.

Consider initial data for a single puncture with mass $M$
at $t=0$,
with $\alpha = 1$, $\beta^i = 0$, $\tilde g_{ij} = \delta_{ij}$,
$\tilde A_{ij} = 0$, $K = 0$, and $\phi_0 = \log(1+\frac{M}{2r})$. When
inserted into the BSSN and gauge equations, the data
evolves and develops certain powers of $r$ at the puncture
\cite{Alcubierre02a}. If a regular, stationary state is reached, then
all variables should possess power series expansions at $r=0$ that satisfy
$\lie_{\beta_{sls}} u + F = 0$.
We make the following ansatz
for the single, non-moving puncture case in spherical symmetry:
\bea
\label{exppsim2}
	\psi^{-2} &=& e^{-2\phi} = p_1 r + p_2 r^2 + O(r^3),
\\
\label{expg}
	\tilde g_{ij} &=& \delta_{ij} + O(r^2),
\\
\label{expA}
	\tilde A_{ij} &=& (A_0 + A_1 r) (\delta_{ij} - 3 n_i n_j) + O(r^2),
\\
\label{expK}
	K &=& K_0 + K_1 r + O(r^2),
\\
\label{expa}
	\alpha &=& a_0 + a_1 r + O(r^2), 
\\
\label{expb}
	\beta^i &=& (b_1 r + b_2 r^2) n^i + O(r^3).
\eea
Here $r=(x^2+y^2+z^2)^{1/2}$ is the coordinate radius of
quasi-isotropic Cartesian coordinates $(x,y,z)$. 
Note that $r$ is continuous but not differentiable,
the radial normal vector $n^i=x^i/r$ is discontinuous at $r=0$, and
$\partial_i O(r^0)= O(r^{-1})$.
The evolution equations and the constraints result in 8 independent
equations for the 10 coefficients
$(p_1,p_2,A_0,A_1,K_0,K_1,a_0,a_1,b_1,b_2)$, with $M$ and $\eta$ free
parameters. The Gamma-freezing shift condition 
does not give an additional condition for a stationary solution. A first result
is that $a_1\neq 0$ and $K_0\neq0$ are required for nontrivial
stationarity; a more regular solution is not consistent with
the equations.  For consistency $a_0=0$, that is the lapse has
collapsed at the puncture. 
Further simple relations are $b_1=2K_0$ and $b_2=-3A_0a_1$ 
from the evolution equations for the lapse and the metric.

\begin{figure}[t]
\centering
\includegraphics[width=40mm]{fig1a}
\quad
\includegraphics[width=40mm,height=40mm]{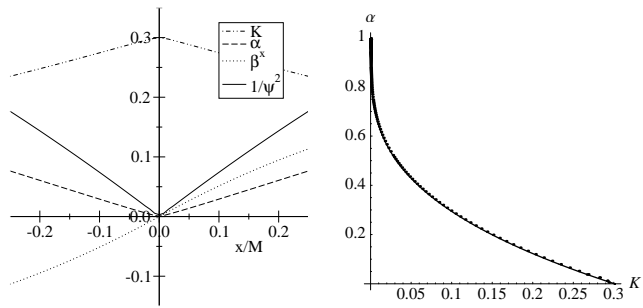}
\caption{Numerical evolutions of a spherically symmetric puncture result in 
specific kinks at the puncture (left). The figure on the right represents a
true numerical experiment. We compare two spherical slices by plotting
$\alpha$ as a function of $K$. The dots are the data from the 3D numerical
evolution, the solid line was obtained by independently integrating up Eq.~(9). 
At the puncture, $\alpha\approx 0.0$ and $K\approx 0.3$.}
\label{fig}
\end{figure}

Therefore, assuming existence of a stationary solution with 
some minimal regularity at the puncture implies specific predictions
for kinks and discontinuities in our variables.
We now compare the ansatz to numerical
results for a
Schwarzschild puncture evolved to $t=50M$, at which time the data
appears to have reached a stationary state to within a few percent in
all variables. The runs were performed with the BAM code, using fixed-mesh
refinement and fourth-order accurate finite-difference stencils 
\cite{Bruegmann:2006at}. For a central resolution of $M/128$ the
run is well within the convergent regime. When zooming into the data at this
resolution, Fig.~\ref{fig}, {\em we do find the kinks and
discontinuities consistent with the ansatz}.  Although closer inspection 
shows some numerical artefacts near the
puncture (in particular in $\tilde A_{ij}$ and the derivatives), they
remain localized and comparatively small, and the
numerical data are approximated well by the ansatz.

\begin{figure}[t]
\centering
\includegraphics[width=40mm]{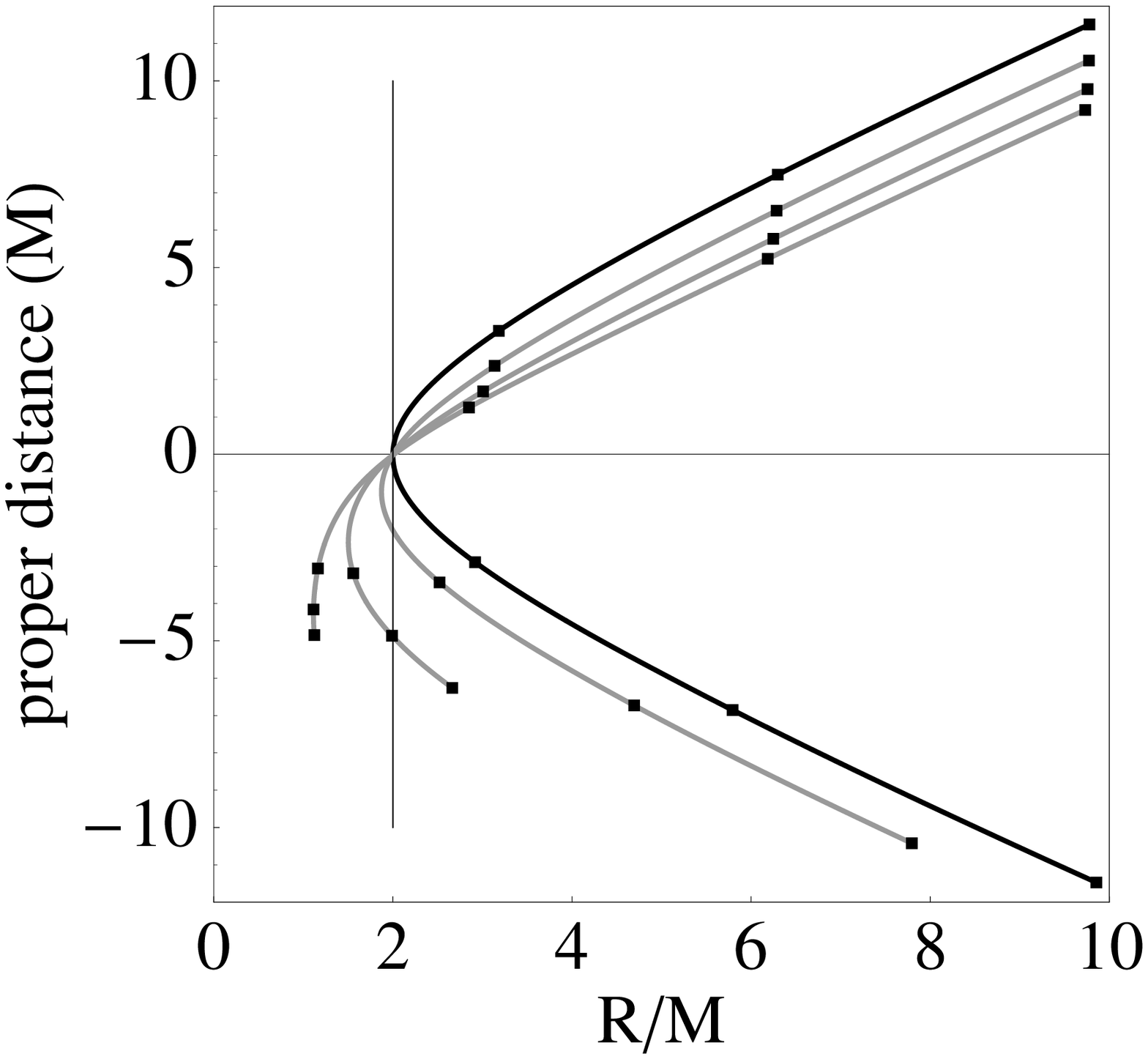}
\quad
\includegraphics[width=40mm]{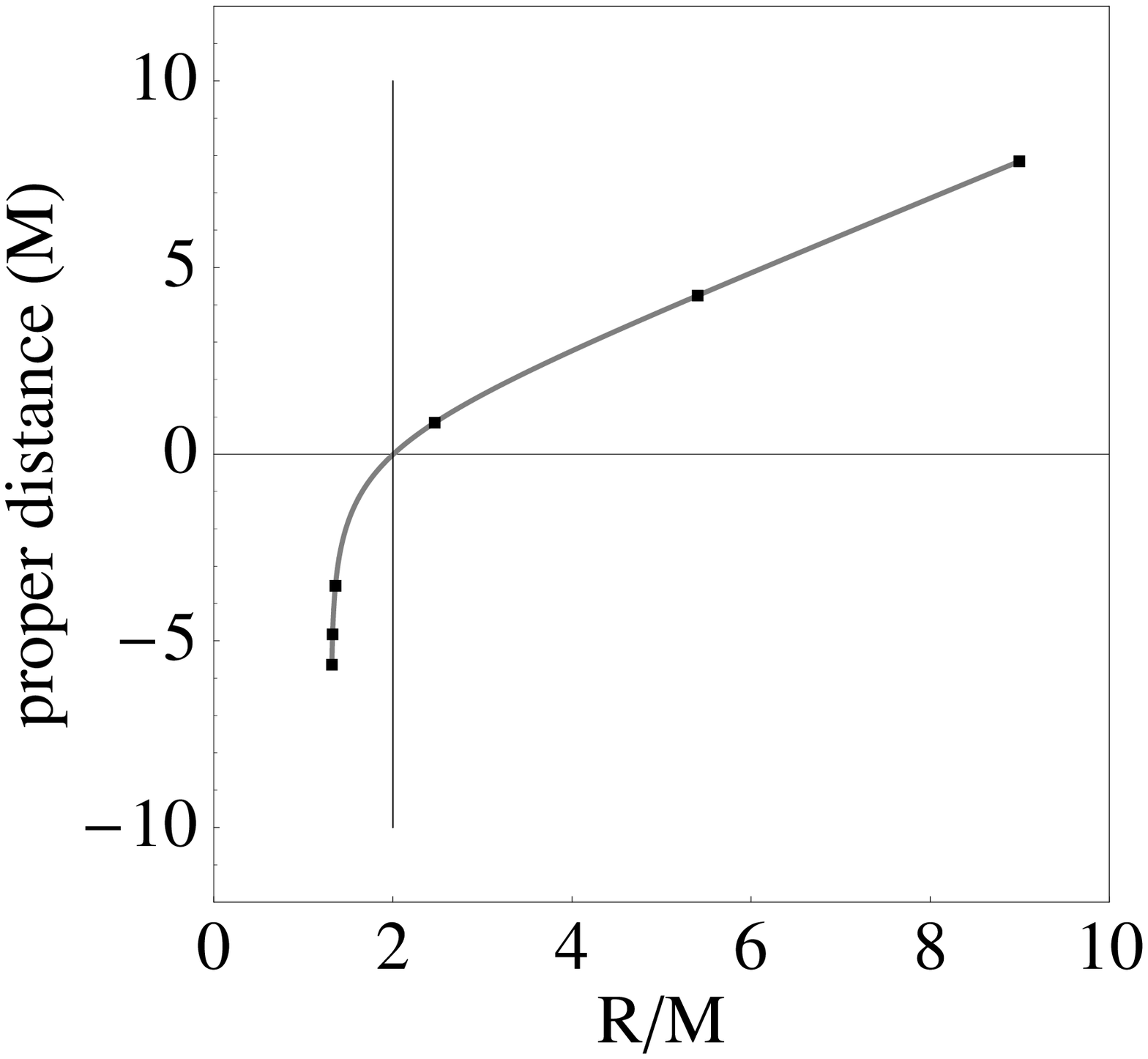}

\caption{Schwarzschild coordinate $R$ versus proper 
distance from the (outer) horizon. The left panel shows the slices at $t = 0,1,2,3M$, 
and the right panel shows the slice at $t = 50M$. The final numerical slice
terminates at $R \approx 1.3M$. The vertical line indicates the horizon at $R = 2M$, and
the six points represent $x/M = 1/40,1/20,1/8,2,5,8$ on each slice.
}
\label{fig:embedding}
\end{figure}

The most intriguing implication of the power series expansion is
that {\em the singularity in the conformal factor changes} during
evolution. With $a_0=0$,
\beq
  \partial_t\phi=\beta^i\partial_i\phi + \frac{1}{6}\partial_i\beta^i +O(r) 
\simeq b_1 r\partial_r\phi + \frac{b_1}{2}.
\eeq
For the conformal factor of the initial data, 
we have $r\partial_r\phi_0\simeq-1$, and 
$\partial_t\phi_0\simeq -b_1/2$. With $b_1\neq0$ we must
conclude that the evolution of the conformal factor will stop only
when $r\partial_r\phi\simeq-1/2$. 
In other words,
\beq
	\psi_0 = O(1/r) \quad \mbox{evolves into} 
    \quad \psi = O(1/\sqrt{r}),
\eeq
which is built into (\ref{exppsim2}).
The puncture still marks a coordinate singularity, but since the areal
radius of the sphere $r=0$ is $R_0 = \lim_{r\rightarrow 0} r\psi^2 = 1/p_1$, 
the slice no longer reaches the other asymptotically flat end of the
Brill-Lindquist wormhole, but ends at a finite areal radius, 
$R_0 \approx 1.3M$, with $p_1$ determined
numerically. Figure~\ref{fig:embedding} illustrates the evolution of a
Schwarzschild black hole. We have plotted Schwarzschild $R$ versus  
proper distance from the horizon. The initial slice connects two
asymptotically flat regions, but during a 1+log/$\Gamma$-driver evolution 
the numerical slice loses contact with the second asymptotically flat end.
It eventually reaches a stationary state and terminates at $R \approx 1.3M$.

%
%

We now show by explicit construction in the proper distance gauge
(which is regular through the horizon), that such stationary slices exist
globally.
Stationary 1+log slices of Schwarzschild are also considered in a
different context in~\cite{Buchman:2005ub}.
The proper distance coordinate is denoted as $l$,
$\partial_l f = f'$, 
$\alpha$ is the Killing lapse,  $\beta$ the Killing radial shift component and
$R$ the areal radius (the Schwarzschild radial coordinate). 
On any spherical slice through the Schwarzschild solution one finds $R' = \alpha$ and $\alpha^2 -
\beta^2 = 1 -2M/R$, and the equation for the stationary 1+log slices of
Schwarzschild, $\beta\alpha' = 2\alpha K$, becomes (using, for example, \cite{NOM06})
\begin{equation}
R'' = \frac{2 R'}{R} \frac{\frac{3M}{R} - 2 + 2R'^2}{\frac{2M}{R} -1 + R'^2 - 2R'}\label{eq:R''}.
\end{equation} It is clear that the right-hand side of  (\ref{eq:R''}) is singular
whenever $2M/R - 1 + R'^2 - 2R'= 0$. It can also be shown that any solution of 
(\ref{eq:R''}) that is suitably asymptotically flat, i.e., $R \approx l$, must pass through
such a singular point. The only way of resolving this difficulty is for the
numerator of (\ref{eq:R''}) to simultaneously vanish at the ``singular'' point. We have two
options: either (a) $R' = 0$, or (b) $3M/R - 2 + 2R'^2 = 0$. Let us discuss them
separately. In the case (a) we have $R' = 0, R = 2M$. Therefore the slice passes through
the bifurcation sphere. This is the standard moment of
time symmetry slice through the Schwarzschild solution, which obviously satisfies the
stationary 1+log equation because $\beta \equiv 0$ and $K \equiv 0$. In case (b), the 
case we are really interested in, we can solve the pair of
simultaneous equations to give $R' = \sqrt{10} - 3, R \approx 1.54 M$. This solution
corresponds to {\em two} slices in the Schwarzschild solution. These are
mirror images of each other, 
one in the upper half plane, one in the lower. These slices do not continue into the
left quadrant of the extended Schwarzschild solution. Rather they asymptote to a cylinder in
the upper (lower) quadrant of fixed radius. This agrees with the numerical observation 
that the singularity in the conformal factor changes from $1/r$
to $1/\sqrt{r}$. These three are the only asymptotically regular solutions of
the stationary 1+log  
equation, up to isometries. It is easy to show that the actual slice exponentially approaches 
a cylinder of radius $R_0$, i.e.,
$R \approx R_0 + A\exp B l$ as $l \rightarrow - \infty$, where 
$$
B = 2 {R_0}^{-1} \left(3 M - 2 R_0\right) \left(2 M - R_0\right)^{-1} .  
$$
The value of the lapse at the horizon, which can be used as a simple
horizon-finding or merger-time criterion, evaluates to $\alpha(R=2M)\approx
0.376$.
It is also possible to produce an algebraic solution in terms of an implicit 
equation for $\alpha$ and $R$; determining $R_0 \approx 1.31241 M$ from
$$
3 \sinh ^{-1}(3) = \log \left(128 (2 - \frac{R_0}{M})\right)+3 \log
\left(\frac{R_0}{M}\right)+\sqrt{10} - 3 .
$$

For comparison we have also studied harmonic and maximal slicing.
With harmonic slicing, 
 (which can be generalized to the Kerr spacetime \cite{Bona88,Cook97b}),
one again finds a moment of time symmetry slice and two mirror copies
of a slice that puncture evolutions driven toward stationarity should approach. 
The slices hit the singularity at $R = 0$, $K$ blows up as $R\rightarrow 0$, 
and at the horizon $\alpha(R = 2M) = 1/2$.
Fixed-puncture evolutions have used 1+log slicing without shift term.
Stationary slices must then be maximal~\cite{Estabrook73,Beig98,Reimann:2003zd},
of the ``odd lapse'' type, $\alpha^2=1 -\frac{2M}{R} + \frac{C^2}{R^4}$.
For $C = 3\sqrt{3}M^2/4$ such slices will again
approach a cylinder, of constant $R_0 = 3 M/2$.

%
%

\paragraph{Discussion. ---}

Moving-puncture evolutions of a Schwarzschild black hole approach a
stationary slice that neither reaches an internal asymptotically flat
end nor hits the physical singularity, as might be expected for a
stationary slice with non-negative lapse \cite{Hannam:2003tv}. Rather,
the slice ends at a throat at finite Schwarzschild radius, but infinite
proper distance from the apparent horizon. This changes the
singularity structure of the ``puncture''. It is still a puncture in
that there is a coordinate singularity at a single point in the
numerical coordinates, but it does not correspond to an asymptotically
flat end. In the course of Schwarzschild evolutions we have found that the
throat does collapse to the origin. Where one would have
expected an inner and an outer horizon, we find only one zero in the
norm of $(\frac{\partial}{\partial t})^a$, corresponding to the outer
horizon. An under-resolved region does develop in the
spacetime (it is the region between the throat and the interior
spacelike infinity), but we are pushed out of causal contact with it. 
The throat itself has receded to infinite proper
distance from the outer horizon. Matter fields or gravitational radiation will be trapped
between the inner horizon and the throat, because unlike the gauge their
propagation is limited by the speed of light; this issue is left to
future work.

The main result about regularity is contained in the specific kinks
and discontinuities discovered in both the power series and the
numerical data, which are implied by our stationary solution. The
construction of an explicit power series solution in $r$ for a single
Schwarzschild puncture can be repeated for a single puncture with
Bowen-York linear momentum.
Preliminary results indicate that
the shift can acquire the constant leading term responsible for the
motion of the punctures, and that there are no new fundamental
regularity issues. The power series ansatz matches nicely the
numerical data for a Schwarzschild puncture.
The fourth-order schemes
(both centered and upwind) in use have been seen to be reasonably
successful in this context, the lack of differentiability not
withstanding \cite{Campanelli:2005dd,Baker05a}.
Incidentally, the introduction of $\chi$ results in
$O(r^2)$ terms and not $O(r^4)$ as expected, but this still results in
significantly cleaner data than the $\log r$ method.
Perhaps most importantly, our analysis suggests a concrete remedy
if the remaining numerical issues at the puncture create problems in
simulations of black-hole binaries, namely to resort to finite
differencing that is expertly adapted to the discontinuities at hand.

Our results suggest several directions for future research directly
relevant to the black-hole binary problem, such as perturbations of
the stationary solutions (including constraint violating perturbations
to check constraint stability of evolution systems), the
clarification of numerical issues at the discontinuities, 
and the construction of initial
data adapted to stationarity, e.g.\ of asymptotically
cylindrical data.  
In the Schwarzschild case local 
properties of the stationary solution allow one to directly read off spacetime 
properties from a numerical solution, e.g., the puncture 
value of $K$ determines the mass; an extension to two moving, spinning black holes
would be very valuable in numerical evolutions.

%
%

We thank Jose Gonzalez, Pedro Marronetti, Uli Sperhake, and Wolfgang
Tichy for discussions.  This work was supported by the DFG grant
SFB/Transregio~7 ``Gravitational Wave Astronomy'', and by computer
time allocations at HLRS Stuttgart and LRZ Munich.

\bibliography{refs}

\begin{thebibliography}{21}
\expandafter\ifx\csname natexlab\endcsname\relax\def\natexlab#1{#1}\fi
\expandafter\ifx\csname bibnamefont\endcsname\relax
  \def\bibnamefont#1{#1}\fi
\expandafter\ifx\csname bibfnamefont\endcsname\relax
  \def\bibfnamefont#1{#1}\fi
\expandafter\ifx\csname citenamefont\endcsname\relax
  \def\citenamefont#1{#1}\fi
\expandafter\ifx\csname url\endcsname\relax
  \def\url#1{\texttt{#1}}\fi
\expandafter\ifx\csname urlprefix\endcsname\relax\def\urlprefix{URL }\fi
\providecommand{\bibinfo}[2]{#2}
\providecommand{\eprint}[2][]{\url{#2}}

\bibitem[{\citenamefont{Pretorius}(2005)}]{Pretorius:2005gq}
\bibinfo{author}{\bibfnamefont{F.}~\bibnamefont{Pretorius}},
  \bibinfo{journal}{Phys. Rev. Lett.} \textbf{\bibinfo{volume}{95}},
  \bibinfo{pages}{121101} (\bibinfo{year}{2005}).

\bibitem[{\citenamefont{Campanelli et~al.}(2006)\citenamefont{Campanelli,
  Lousto, Marronetti, and Zlochower}}]{Campanelli:2005dd}
\bibinfo{author}{\bibfnamefont{M.}~\bibnamefont{Campanelli}},
  \bibinfo{author}{\bibfnamefont{C.~O.} \bibnamefont{Lousto}},
  \bibinfo{author}{\bibfnamefont{P.}~\bibnamefont{Marronetti}},
  \bibnamefont{and}
  \bibinfo{author}{\bibfnamefont{Y.}~\bibnamefont{Zlochower}},
  \bibinfo{journal}{Phys. Rev. Letter} \textbf{\bibinfo{volume}{96}},
  \bibinfo{pages}{111101} (\bibinfo{year}{2006}).

\bibitem[{\citenamefont{Baker et~al.}(2006)\citenamefont{Baker, Centrella,
  Choi, Koppitz, and van Meter}}]{Baker05a}
\bibinfo{author}{\bibfnamefont{J.~G.} \bibnamefont{Baker}},
  \bibinfo{author}{\bibfnamefont{J.}~\bibnamefont{Centrella}},
  \bibinfo{author}{\bibfnamefont{D.-I.} \bibnamefont{Choi}},
  \bibinfo{author}{\bibfnamefont{M.}~\bibnamefont{Koppitz}}, \bibnamefont{and}
  \bibinfo{author}{\bibfnamefont{J.}~\bibnamefont{van Meter}},
  \bibinfo{journal}{Phys. Rev. Lett.} \textbf{\bibinfo{volume}{96}},
  \bibinfo{pages}{111102} (\bibinfo{year}{2006}).

\bibitem[{\citenamefont{Br\"ugmann et~al.}(2004)\citenamefont{Br\"ugmann,
  Tichy, and Jansen}}]{Bruegmann:2003aw}
\bibinfo{author}{\bibfnamefont{B.}~\bibnamefont{Br\"ugmann}},
  \bibinfo{author}{\bibfnamefont{W.}~\bibnamefont{Tichy}}, \bibnamefont{and}
  \bibinfo{author}{\bibfnamefont{N.}~\bibnamefont{Jansen}},
  \bibinfo{journal}{Phys. Rev. Lett.} \textbf{\bibinfo{volume}{92}},
  \bibinfo{pages}{211101} (\bibinfo{year}{2004}).

\bibitem[{\citenamefont{Brill and Lindquist}(1963)}]{Brill63}
\bibinfo{author}{\bibfnamefont{D.~} \bibnamefont{Brill}} \bibnamefont{and}
  \bibinfo{author}{\bibfnamefont{R.~} \bibnamefont{Lindquist}},
  \bibinfo{journal}{Phys. Rev.} \textbf{\bibinfo{volume}{131}},
  \bibinfo{pages}{471} (\bibinfo{year}{1963}).

\bibitem[{\citenamefont{Brandt and Br{\"u}gmann}(1997)}]{Brandt97b}
\bibinfo{author}{\bibfnamefont{S.}~\bibnamefont{Brandt}} \bibnamefont{and}
  \bibinfo{author}{\bibfnamefont{B.}~\bibnamefont{Br{\"u}gmann}},
  \bibinfo{journal}{Phys. Rev. Lett.} \textbf{\bibinfo{volume}{78}},
  \bibinfo{pages}{3606} (\bibinfo{year}{1997}).

\bibitem[{\citenamefont{Alcubierre et~al.}(2001)\citenamefont{Alcubierre,
  Benger, Br\"ugmann, Lanfermann, Nerger, Seidel, and
  Takahashi}}]{Alcubierre00b}
\bibinfo{author}{\bibfnamefont{M.}~\bibnamefont{Alcubierre}},
  \bibinfo{author}{\bibfnamefont{W.}~\bibnamefont{Benger}},
  \bibinfo{author}{\bibfnamefont{B.}~\bibnamefont{Br\"ugmann}},
  \bibinfo{author}{\bibfnamefont{G.}~\bibnamefont{Lanfermann}},
  \bibinfo{author}{\bibfnamefont{L.}~\bibnamefont{Nerger}},
  \bibinfo{author}{\bibfnamefont{E.}~\bibnamefont{Seidel}}, \bibnamefont{and}
  \bibinfo{author}{\bibfnamefont{R.}~\bibnamefont{Takahashi}},
  \bibinfo{journal}{Phys. Rev. Lett.} \textbf{\bibinfo{volume}{87}},
  \bibinfo{pages}{271103} (\bibinfo{year}{2001}).

\bibitem[{\citenamefont{Br{\"u}gmann}(1999)}]{Bruegmann97}
\bibinfo{author}{\bibfnamefont{B.}~\bibnamefont{Br{\"u}gmann}},
  \bibinfo{journal}{Int. J. Mod. Phys. D} \textbf{\bibinfo{volume}{8}},
  \bibinfo{pages}{85} (\bibinfo{year}{1999}).

\bibitem[{\citenamefont{Diener et~al.}(2006)}]{Diener:2005mg}
\bibinfo{author}{\bibfnamefont{P.}~\bibnamefont{Diener}} \bibnamefont{et~al.},
  \bibinfo{journal}{Phys. Rev. Lett.} \textbf{\bibinfo{volume}{96}},
  \bibinfo{pages}{121101} (\bibinfo{year}{2006}).

\bibitem[{\citenamefont{Shibata and Nakamura}(1995)}]{Shibata95}
\bibinfo{author}{\bibfnamefont{M.}~\bibnamefont{Shibata}} \bibnamefont{and}
  \bibinfo{author}{\bibfnamefont{T.}~\bibnamefont{Nakamura}},
  \bibinfo{journal}{Phys. Rev. D} \textbf{\bibinfo{volume}{52}},
  \bibinfo{pages}{5428} (\bibinfo{year}{1995}).

\bibitem[{\citenamefont{Baumgarte and Shapiro}(1999)}]{Baumgarte99}
\bibinfo{author}{\bibfnamefont{T.~W.} \bibnamefont{Baumgarte}}
  \bibnamefont{and} \bibinfo{author}{\bibfnamefont{S.~L.}
  \bibnamefont{Shapiro}}, \bibinfo{journal}{Phys. Rev. D}
  \textbf{\bibinfo{volume}{59}}, \bibinfo{pages}{024007}
  (\bibinfo{year}{1999}).

\bibitem[{\citenamefont{Alcubierre et~al.}(2003)\citenamefont{Alcubierre,
  Br\"ugmann, Diener, Koppitz, Pollney, Seidel, and Takahashi}}]{Alcubierre02a}
\bibinfo{author}{\bibfnamefont{M.}~\bibnamefont{Alcubierre}},
  \bibinfo{author}{\bibfnamefont{B.}~\bibnamefont{Br\"ugmann}},
  \bibinfo{author}{\bibfnamefont{P.}~\bibnamefont{Diener}},
  \bibinfo{author}{\bibfnamefont{M.}~\bibnamefont{Koppitz}},
  \bibinfo{author}{\bibfnamefont{D.}~\bibnamefont{Pollney}},
  \bibinfo{author}{\bibfnamefont{E.}~\bibnamefont{Seidel}}, \bibnamefont{and}
  \bibinfo{author}{\bibfnamefont{R.}~\bibnamefont{Takahashi}},
  \bibinfo{journal}{Phys. Rev. D} \textbf{\bibinfo{volume}{67}},
  \bibinfo{pages}{084023} (\bibinfo{year}{2003}).

\bibitem[{\citenamefont{van Meter et~al.}(2006)\citenamefont{van Meter, Baker,
  Koppitz, and Choi}}]{vanMeter:2006vi}
\bibinfo{author}{\bibfnamefont{J.~R.} \bibnamefont{van Meter}},
  \bibinfo{author}{\bibfnamefont{J.~G.} \bibnamefont{Baker}},
  \bibinfo{author}{\bibfnamefont{M.}~\bibnamefont{Koppitz}}, \bibnamefont{and}
  \bibinfo{author}{\bibfnamefont{D.-I.} \bibnamefont{Choi}},
  \bibinfo{journal}{Phys. Rev.} \textbf{\bibinfo{volume}{D73}},
  \bibinfo{pages}{124011} (\bibinfo{year}{2006}).

\bibitem[{\citenamefont{Gundlach and Martin-Garcia}(2006)}]{Gundlach:2006tw}
\bibinfo{author}{\bibfnamefont{C.}~\bibnamefont{Gundlach}} \bibnamefont{and}
  \bibinfo{author}{\bibfnamefont{J.~M.} \bibnamefont{Martin-Garcia}},
  \bibinfo{journal}{Phys. Rev.} \textbf{\bibinfo{volume}{D74}},
  \bibinfo{pages}{024016} (\bibinfo{year}{2006}).

\bibitem[{\citenamefont{Br{\"u}gmann et~al.}(2006)\citenamefont{Br{\"u}gmann,
  Gonz\'alez, Hannam, Husa, Sperhake, and Tichy}}]{Bruegmann:2006at}
\bibinfo{author}{\bibfnamefont{B.}~\bibnamefont{Br{\"u}gmann}},
  \bibinfo{author}{\bibfnamefont{J.~A.} \bibnamefont{Gonz\'alez}},
  \bibinfo{author}{\bibfnamefont{M.}~\bibnamefont{Hannam}},
  \bibinfo{author}{\bibfnamefont{S.}~\bibnamefont{Husa}},
  \bibinfo{author}{\bibfnamefont{U.}~\bibnamefont{Sperhake}}, \bibnamefont{and}
  \bibinfo{author}{\bibfnamefont{W.}~\bibnamefont{Tichy}}
  (\bibinfo{year}{2006}), \eprint{gr-qc/0610128}.

\bibitem[{\citenamefont{Buchman and Bardeen}(2005)}]{Buchman:2005ub}
\bibinfo{author}{\bibfnamefont{L.~T.} \bibnamefont{Buchman}} \bibnamefont{and}
  \bibinfo{author}{\bibfnamefont{J.~M.} \bibnamefont{Bardeen}},
  \bibinfo{journal}{Phys. Rev.} \textbf{\bibinfo{volume}{D72}},
  \bibinfo{pages}{124014} (\bibinfo{year}{2005}).

 \bibitem[{\citenamefont{N. \'O Murchadha and K. Roszkowski}(2008)}]{NOM06}
 \bibinfo{author}{\bibfnamefont{N.}~\bibnamefont{\'O Murchadha}} \bibnamefont{and}
   \bibinfo{author}{\bibfnamefont{K.}~\bibnamefont{Roszkowski}},
   \bibinfo{journal}{Class. Quantum Grav.} \textbf{\bibinfo{volume}{23}},
   \bibinfo{pages}{539} (\bibinfo{year}{2006}).

 \bibitem[{\citenamefont{Bona and Mass\'{o}}(1988)}]{Bona88}
 \bibinfo{author}{\bibfnamefont{C.}~\bibnamefont{Bona}} \bibnamefont{and}
   \bibinfo{author}{\bibfnamefont{J.}~\bibnamefont{Mass\'{o}}},
   \bibinfo{journal}{Phys. Rev. D} \textbf{\bibinfo{volume}{38}},
   \bibinfo{pages}{2419} (\bibinfo{year}{1988}).

 \bibitem[{\citenamefont{Cook and Scheel}(1997)}]{Cook97b}
 \bibinfo{author}{\bibfnamefont{G.~B.} \bibnamefont{Cook}} \bibnamefont{and}
   \bibinfo{author}{\bibfnamefont{M.~A.} \bibnamefont{Scheel}},
   \bibinfo{journal}{Phys. Rev. D} \textbf{\bibinfo{volume}{56}},
   \bibinfo{pages}{4775} (\bibinfo{year}{1997}).

\bibitem[{\citenamefont{Estabrook et~al.}(1973)\citenamefont{Estabrook,
  Wahlquist, Christensen, DeWitt, Smarr, and Tsiang}}]{Estabrook73}
\bibinfo{author}{\bibfnamefont{F.}~\bibnamefont{Estabrook}},
  \bibinfo{author}{\bibfnamefont{H.}~\bibnamefont{Wahlquist}},
  \bibinfo{author}{\bibfnamefont{S.}~\bibnamefont{Christensen}},
  \bibinfo{author}{\bibfnamefont{B.}~\bibnamefont{DeWitt}},
  \bibinfo{author}{\bibfnamefont{L.}~\bibnamefont{Smarr}}, \bibnamefont{and}
  \bibinfo{author}{\bibfnamefont{E.}~\bibnamefont{Tsiang}},
  \bibinfo{journal}{Phys. Rev. D} \textbf{\bibinfo{volume}{7}},
  \bibinfo{pages}{2814} (\bibinfo{year}{1973}).

\bibitem[{\citenamefont{Beig and {\'O~Murchadha}}(1998)}]{Beig98}
\bibinfo{author}{\bibfnamefont{R.}~\bibnamefont{Beig}} \bibnamefont{and}
  \bibinfo{author}{\bibfnamefont{N.}~\bibnamefont{{\'O~Murchadha}}},
  \bibinfo{journal}{Phys. Rev. D} \textbf{\bibinfo{volume}{57}},
  \bibinfo{pages}{4728} (\bibinfo{year}{1998}).

\bibitem[{\citenamefont{Reimann and Br{\"u}gmann}(2004)}]{Reimann:2003zd}
\bibinfo{author}{\bibfnamefont{B.}~\bibnamefont{Reimann}} \bibnamefont{and}
  \bibinfo{author}{\bibfnamefont{B.}~\bibnamefont{Br{\"u}gmann}},
  \bibinfo{journal}{Phys. Rev. D} \textbf{\bibinfo{volume}{69}},
  \bibinfo{pages}{044006} (\bibinfo{year}{2004}).

\bibitem[{\citenamefont{Hannam et~al.}(2003)\citenamefont{Hannam, Evans, Cook,
  and Baumgarte}}]{Hannam:2003tv}
\bibinfo{author}{\bibfnamefont{M.~D.} \bibnamefont{Hannam}},
  \bibinfo{author}{\bibfnamefont{C.~R.} \bibnamefont{Evans}},
  \bibinfo{author}{\bibfnamefont{G.~B.} \bibnamefont{Cook}}, \bibnamefont{and}
  \bibinfo{author}{\bibfnamefont{T.~W.} \bibnamefont{Baumgarte}},
  \bibinfo{journal}{Phys. Rev. D} \textbf{\bibinfo{volume}{68}},
  \bibinfo{pages}{064003} (\bibinfo{year}{2003}).

\end{thebibliography}

\end{document}